\documentclass[11pt]{article}
\title{Optimizing the Generation and Transmission Capacity of Offshore Wind Parks under Weather Uncertainty}
\author{David Kröger\footnote{Corresponding author. \href{mailto:david.kroeger@tu-dortmund.de}{david.kroeger@tu-dortmund.de}}, Jan Peper, Nils Offermann, Christian Rehtanz  \\
	Institute of Energy Systems, Energy Efficiency and Energy Economics \\
	TU Dortmund University, Dortmund, Germany
}

\usepackage[backend=bibtex,style=numeric,sorting=none]{biblatex}
\usepackage[a4paper, left=3.5cm, right=4.5cm, top=3cm]{geometry} 
\usepackage{xurl}
\usepackage{amsmath}
\usepackage{hyperref}
\usepackage{framed}
\usepackage[most]{tcolorbox}
\usepackage{booktabs,threeparttable} 
\usepackage{subcaption}

\date{2023}

\begin{document}
									
\maketitle
				
\begin{abstract}
Offshore wind power in the North Sea is considered a main pillar in Europe’s future energy system. A key challenge lies in determining the optimal spatial capacity allocation of offshore wind parks in combination with the dimensioning and layout of the connecting high-voltage direct current grid infrastructure. To determine economically cost optimal configurations, we apply an integrated capacity and transmission expansion problem within a pan-European electricity market and transmission grid model with a high spatial and temporal granularity. By conducting scenario analysis for the year 2030 with a gradually increasing CO\textsubscript{2} price, possible offshore expansion paths are derived and presented. Special emphasis is laid on the effects of weather uncertainty by incorporating data from 21 historical weather years in the analysis. Two key findings are (i) an expansion in addition to the existing offshore wind capacity of 0\,GW (136\,\texteuro/tCO\textsubscript{2}), 12\,GW (159\,\texteuro/tCO\textsubscript{2}) and 30\,GW (186\,\texteuro/tCO\textsubscript{2}) dependent on the underlying CO\textsubscript{2} price. (ii) A strong sensitivity of the results towards the underlying weather data highlighting the importance of incorporating multiple weather years.

\end{abstract}

\newpage
							
\section{Offshore Wind Power in the North Sea}\label{sec1} 
The North Sea is a shelf sea in northwestern Europe that is connected to the Atlantic in the south and the Norwegian Sea in the north \cite{Schwarzer.2019}. In total, the North Sea covers an area of approximately 570.000 km\textsuperscript{2} with an average water depth of 94\,m \cite{Schwarzer.2019}. However, in the areas south of the axis \textit{Scarborough}, \textit{Doggerbank} and \textit{Jütlandbank}, the water depth is generally less than 50\,m \cite{Schwarzer.2019,karteNordseeWiki}. Along the coastal regions of Great Britain, France, Belgium, Netherlands, Germany, Denmark, Sweden and Norway around 80 million people live which makes the North Sea an exceptionally important economic area \cite{KristianUhlenbrock.2010}. Overall, the shallow waters, the close proximity to population centers in Europe as well as high wind availability make the North Sea an outstanding area for future offshore wind power production. 

Renewable energy source (RES) production from offshore wind turbines began in the early 1990s in Denmark, Sweden and Netherlands \cite{KristianUhlenbrock.2010}. Since then, the total installed capacity of offshore wind turbines in the North Sea has been expanded to almost 20\,GW (2020) which constitutes 78\,\% of the total wind offshore capacity in Europe \cite{WindEuropeStatistics}. Due to techno-economical, regulatory, ecological and safety aspects, only a fraction of the total area covered by the North Sea is viable for the further development of offshore wind parks (OWPs). Besides already existing oil,  natural gas, electricity and communication infrastructures, other commercial ventures such as sand and gravel mining, fishing, merchant shipping and tourism further restrict the remaining available area for OWPs \cite{Gusatu.2020}. Besides the commercial usage sites, there are dedicated military realms and environmental protection areas \cite{Gusatu.2020} that reduce the available space for OWPs even more. 
			
An intriguing research question from an overall systemic point of view is the maximization of welfare in Europe considering the optimal capacity allocation of OWPs and their corresponding high-voltage direct current (HVDC)\footnote{The advantages of using HVDC are elaborated in Section \ref{sec2}.} grid layout taking into account restrictions from the coupled European electricity markets and the continental transmission grid. The arising optimization problem can be classified as a combined generation and transmission capacity expansion (GTCE) problem in which both the total capacity of each OWP and the capacity of the corresponding grid connection between offshore nodes amongst themselves and onshore nodes is determined. 

We contribute to the current discussion by presenting optimal OWP capacity and grid topology configurations and discuss their impact on the pan-European electricity markets. Therefore, endogenous modeling of both OWP and transmission capacities is applied considering a high level of spatial and temporal detail using clustering algorithms for complexity reduction. The models are implemented in an expansion problem formulation within the pan-European electricity market and transmission grid model \textit{MILES} \cite{MILES} and applied to the scenario year 2030. Special emphasis is laid on the effects of incorporating different weather years which proves to be a major influencing factor in the determination of an optimal capacity and topology configuration. In summary, the main contributions of this paper are as follows. 

\begin{enumerate}
	\item Endogenous optimization of both the capacity of OWPs as well as the corresponding grid topology considering a high spatial and temporal level of detail. Therefore, clustering approaches in both the temporal and spatial domains are applied to keep the complexity of the optimization problem in check. 
	
	\item By applying an integrated multi-year weather approach, robust optimal capacity and topology configurations are obtained. The multi-year approach is compared against the commonly used single-year approach to illustrate the effects and necessity of incorporating multiple weather years. 
\end{enumerate}

\subsection{State of the Art}
Literature on this research problem can be classified depending on the approach and whether decision-relevant parameters are modeled as endogenous or exogenous variables. In principle, free variables such as OWP and transmission capacity can be modeled endogenously as decision variables or be modeled exogenously by setting the variable to a predefined value and subsequently run a \textit{mere} simulation or reduced optimization. There also exist heuristic approaches in which exogenous parameter combinations are systematically tested out and the best configuration is chosen. Generally, the optimization problem gets more complex with an increasing number of free, independent parameters and the determination of the right balance between complexity and accuracy of results is a challenge in itself. Thus, in the literature different approaches have evolved ranging from full exogenous modeling i.e. full sets of predetermined parameters \cite{Jansen.2022, Egerer.2013, Tosatto.2022, Konstantelos.2017, Thommessen.2021, Roussanaly.2019} over a mixture of both fixed and free parameters \cite{Fliegner.2022, Hentschel.June2019, Trotscher.2011, Svendsen.2013} to pure endogenous modeling \cite{Koivisto.2020, GeaBermudez.2020}. 

\cite{Jansen.2022} conducts a comparison of the \textit{North Sea Wind Power Hub} concept against conventional point-to-point connections for the scenario year 2030 using a pan-European electricity market model. The study finds that if more than 10\,GW of wind is built, the Power hub concept is economically favorable due to the avoided cost for multiple converter platforms in the conventional approach. Furthermore, the study highlights that distribution effects resulting from wind power integration are not felt equally across Europe leading to the creation of winners and losers. 
\cite{Egerer.2013} investigates the pan-European infrastructure project \textit{North and Baltic Seas Grid} using the pan-European electricity market model \textit{ELMOD}. Different connection scenarios are tested consisting of a trade scenario with bilateral contracts and point-to-point connection and another scenario with meshed networks. Results show the superiority of the meshed network in terms of overall welfare gains. Again, this study points out that there can be losers from a distributional perspective in a more connected pan-European electricity market, namely generating firms in France, Germany and Poland, as well as consumers in low-price countries. Furthermore, the study highlights the strong interdependencies between an offshore grid expansion and the subsequent onshore network.
In \cite{Tosatto.2022} an analysis of offshore hubs in the North Sea on the European power system and electricity market is carried out. The study shows that energy hubs in the North Sea contribute to an increase in social welfare in Europe, although the benefits are not shared equally across countries. The complete modeling framework including code and datasets has been made available as an open-access resource. 
\cite{Fliegner.2022} presents a method to optimize offshore grid topologies connecting multiple wind farms and countries. Therefore, a GIS system is used to cluster wind farms and create a permissive graph topology which is then used in a downstream market model to determine optimal investment into new lines based on the GIS created graph. The model is applied to the Baltic Sea Region for the target year 2040. Findings suggest that offshore topologies benefit from bundled transmission paths and clustered wind farms. Furthermore, the study highlights the impacts on the topology results resulting from the level of detail of the modeling. 
\cite{Trotscher.2011} represents an early contribution on the subject of optimal offshore grid expansion using an electricity market model with endogenous modeling of the grid expansion. The developed method is applied in a case study of the North Sea region for a 2030 scenario considering six price areas.
\cite{Koivisto.2020} analyses the optimal transmission and generation investments in the North Sea region towards 2050 using the energy system model \textit{Balmorel}. The study compares a project-based scenario in which each offshore wind farm is connected individually against an integrated optimization approach in which transmission and generation is optimized together. Findings show that the integrated solutions leads to an overall cost minimum and that a mixture of radial lines and the utilization of transmission infrastructure via the wind hubs provides the optimal interconnection between different countries. 
\cite{GeaBermudez.2020} also investigates optimal offshore transmission grid expansion investments for the integration of large shares of renewable energy in the North Sea region until 2050 with special emphasis on the applied planning horizon. Therefore, an investment optimization of both generation and transmission capacities for different scenarios is applied. The results show that an integrated offshore grid configuration planned over a long planning horizon leads to cost minimization compared to myopic approach. 
Besides the consideration of a pure electric offshore grid, there are other concepts that incorporate the usage of gas infrastructure or include the option of carbon capture and storage. For example, \cite{Thommessen.2021} analyses energy hub systems in the North Sea considering the production and transportation of hydrogen and ammonia. \cite{Roussanaly.2019} investigates the concept of offshore power generation from natural gas with downstream carbon capture and storage to reduce the climate impact. The study finds that the concept offers significant potential for the cost-efficient decarbonisation of the offshore oil and gas industry while its effects on the mainland electricity markets are limited.
			
\section{Grid Integration of Offshore Wind Power}\label{sec2}
\begin{figure}
	\centering
	\includegraphics[width=\textwidth]{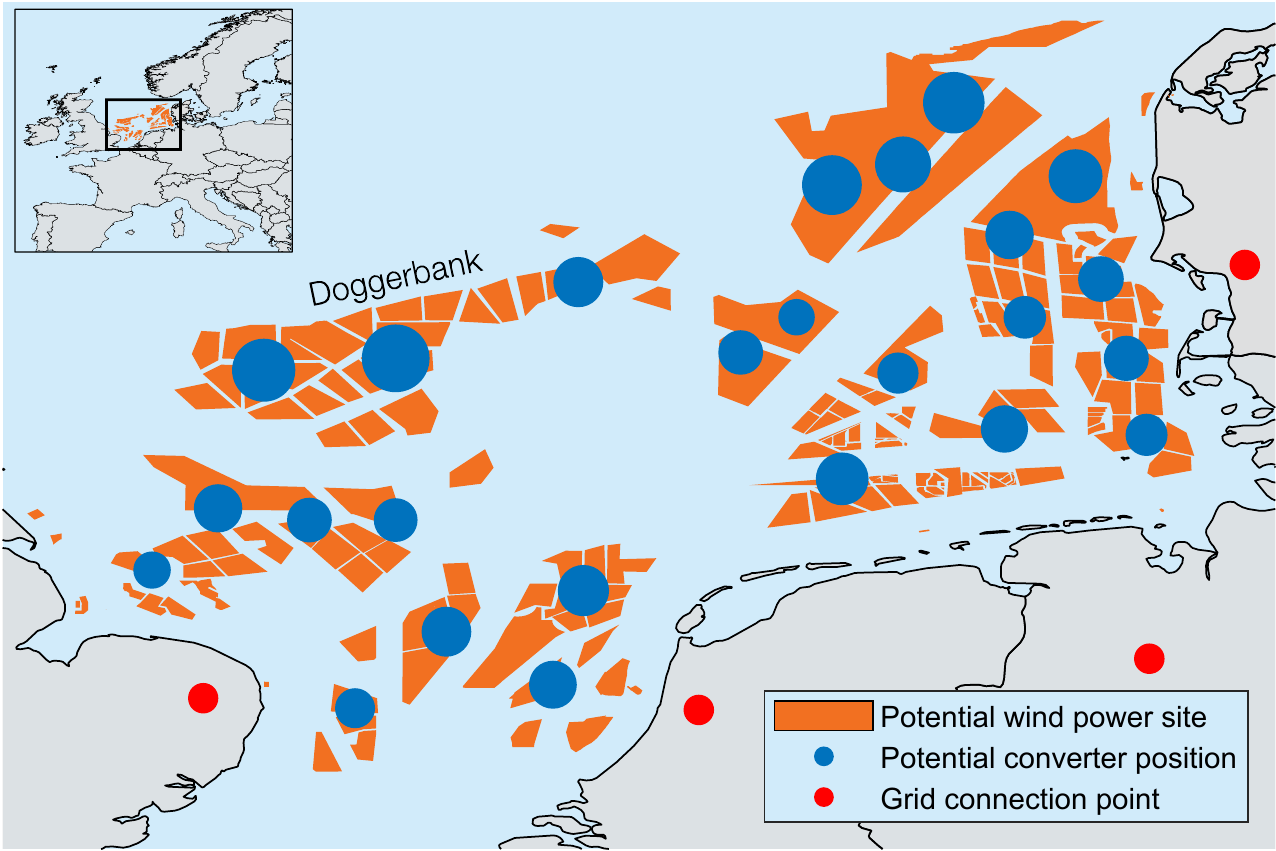}
	\caption{Map of the southern North Sea with potential wind power sites, converter positions and mainland grid connection points. The size of the converter position markers is equivalent to the available wind power pooled to each converter. The spatial clustering algorithm achieved a maximum distance between each point of the superimposed virtual grid of the orange potential sites to the nearest converter site of approx. 70\,km. The proportions of the blue and red markers are equivalent to the available maximum power.}
\label{fig:PotentialflächenClustered} 
\end{figure}

In order to derive the total available area for OWPs in the North Sea, which serves as a fundamental input of the downstream optimization, different data sources are combined. In a study on behalf of the North Sea Wind Power Hub Consortium (NSWPH), which consists of transmission system operators for both gas and electricity - \textit{Energinet}, \textit{TenneT} and \textit{Gasunie} - it has been calculated that approximately 3\,\% of the North Sea with a depth of less than 55 meter remains available for OWPs (14.000\,km\textsuperscript{2}) excluding areas with already other land usages  \cite{AndrevanKuijk.2019}. The study considers planned OWP capacity of 55\,GWs until 2030 and projects an additional 47 - 84\,GWs that can be installed in the North Sea after 2030 depending on the power density applied \cite{AndrevanKuijk.2019}. Further indications of the available space for future OWPs can be found in the 2014 created database \cite{EuropeanMarineObservationandDataNetwork.2023} of the European Marine Observation and Data Network (EMODnet). The database is regularly updated and compiles information such as technical aspects and status on already existing or approved OWPs as well as areas of planned OWPs. Using surface layering and intersection in QGIS \cite{QGIS_software}, data from the NSWPH study and the EMODnet database are combined to represent the future available potential area for OWPs  (Figure \ref{fig:PotentialflächenClustered}). 

Depending on the actual design of the grid connection concept, wind turbines are connected to either a transformer or a converter platform \cite{NEP.2035}. At the transformer respectively converter platform, the wind turbines are pooled and coupled to the mainland grid via an AC or (HV)DC connection \cite{NEP.2035}. A DC connection is generally regarded as technically and economically favourable from a distance of 100\,km and onwards due to the lower losses and larger transfer capacities as well as the non-necessity for a power factor correction \cite{Elliott.2016}. Therefore, most planned and approved projects covering longer distances are planned as a DC grid connection system with decentral transformer platforms to pool the feed-in from wind turbines \cite{Decker.2011}. Using these HVDC links, multiple converter stations can also be connected to one another using a multiterminal approach \cite{Adam.2016} enabling the power transfer between converter stations and the connected mainland grids. Two main advantages of this concept are (i) the potential increase in socio-economic welfare trough the strengthened interconnection between international electricity markets and (ii) an increased redundancy of the grid connection of the OWPs \cite{Chondrogiannis.2016}.

In order to obtain potential locations for the installation of offshore wind turbines, a clustering algorithm is applied on the designated wind potential sites. This is done by superimposing a grid with equidistant nodes to the given regions and the subsequent application of a k-medoids algorithm to all nodes that are located in one of the designated sites. In this process, the number of clusters is increased iteratively until a specified maximal distance from each computed medoid location to its assigned nodes is met. This maximal distance can be based on the grid topology, i.e. the upper limit of the AC-connection of single wind farms to the central HVDC converter. In Figure \ref{fig:PotentialflächenClustered} the position of the potential central HVDC converter sites are marked in blue and of the mainland connection points in red.

\subsection{Temporal Clustering}\label{sec3}
\begin{figure}
	\centering
	\includegraphics[width=\textwidth]{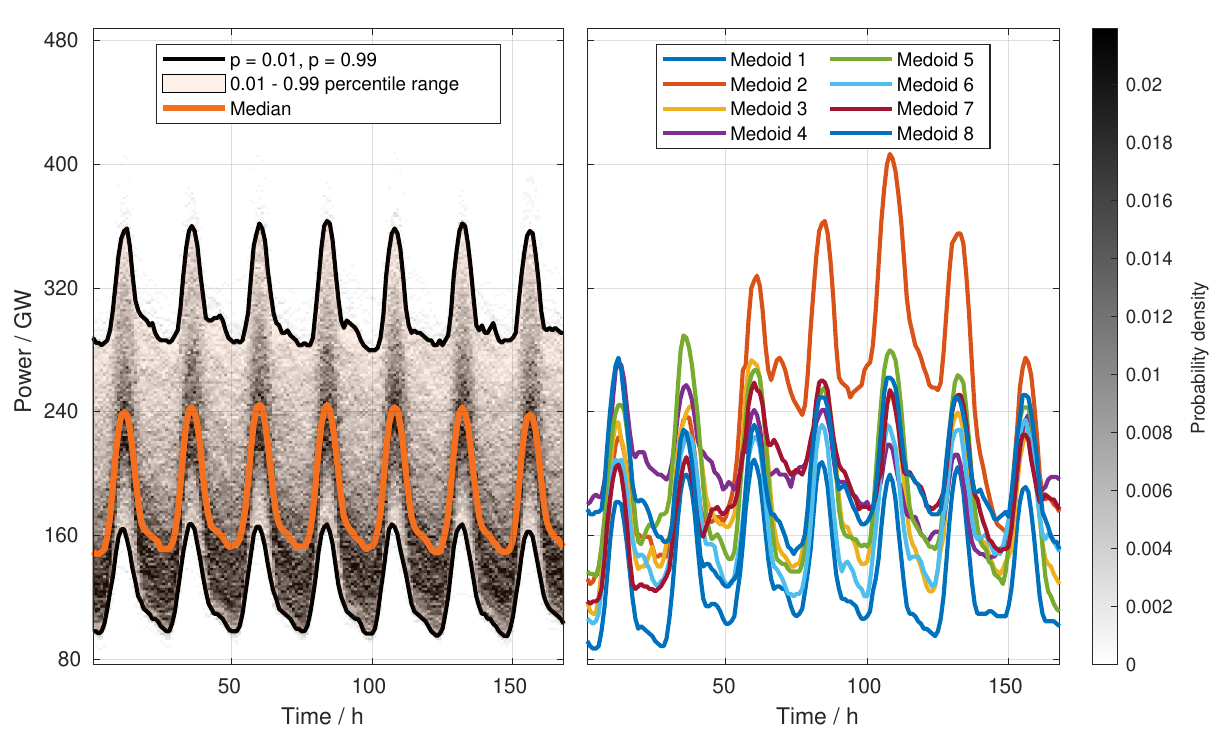}
	\caption{Left: Probability distribution of available RES feed-in of European countries for the target year 2030 based on 21 historical weather years from 1996 - 2017 including the median and the 0.01 and 0.99 percentile range. Right: The eight representative weeks. Medoids 1 and 2 are the weeks with the minimum and maximum occuring values over the 21 years. The occurrence of the weeks in the one-year dispatch period duration are  $N_z = [ 2, 3, 6, 11, 5, 14, 4, 7 ]$. The data has been generated using the pan-European energy system analysis model \textit{MILES} \cite{MILES}.} 
	\label{fig:zeitlichesClustering} 
\end{figure}

The investment optimization is conducted based on the annuity method using a period duration of one year. In order to determine the operational expenditures of the total system during each one-year period, an integrated economic dispatch problem in hourly resolution is solved as part of the optimization problem. Due to the high complexity of the GTCE problem resulting mainly from the interval length of one year, hourly resolution and time-coupling constraints, the underlying input data is clustered into a number of representative weeks and their corresponding occurrence to represent the one-year period. Considering the asset lifetime of OWPs, transmission lines and converters, the underlying input data, especially time series of available RES feed-in and offshore wind power capacity factors, should ideally feature not only the representative behavior of one year, but over their complete lifetime. The use of a single weather year would imply its reoccurrence over the whole assets lifetime. Thus, 21 available historical weather years provided by the German Federal meteorological office DWD\footnote{The weather data is provided in hourly resolution and for whole Europe within the project COSMO-REA6.} \cite{Bollmeyer.2015} are temporally clustered into a number of representative weeks and their corresponding occurrence to account for the one-year period considered in the investment optimization. Instead of directly using the weather data as input for the clustering algorithm, time series of available RES feed-in from all European countries are used due to the nonlinear effects of certain system characteristics for example in case of wind power turbines. Figure \ref{fig:zeitlichesClustering} shows the distribution of the aggregated available RES feed-in for Europe for the underlying complete 21 historical weather years as well as the representative candidates resulting from the temporal clustering. Eight representative candidates have been determined using the nRMSE between the original and clustered data and a convergence criteria of 10\,\%. Furthermore, the clustering algorithm has been modified to incorporate the weeks with the overall minimum and maximum values as representative candidates in any case.

\subsection{Costs and further Scenario Parameters}
The total cost for the construction and operation of the OWPs and the corresponding offshore grid connections comprise of different cost elements. One major cost driver are converters (including auxiliary power electronics) $c^{\mathrm{c,varP,ac-dc}}$ that link the AC and DC grids. In case of a HVDC node-to-node connection (multiterminal), a reduced amount of power electronics are necessary and the costs $c^{\mathrm{c,varP,dc-dc}}$ are mainly driven by direct current circuit breakers. In accordance to \cite{Nieradzinska.2016,cigre.2013} the costs for HVDC multitermals are estimated with 1/6 of the costs for a AC - DC grid connection. Furthermore, a fixed cost term $c^{\mathrm{c,fix}}$ for the erection and construction of sea platforms for the power electronics is considered. The costs for transmission are split up into routing and cable costs, i.e. a small fixed amount for the determination of a corridor $c^{\mathrm{b,fix}}$, a length-dependent part/share for planning procedures and installation onshore $c^{\mathrm{b,on,varL}}$ and offshore $c^{\mathrm{b,off,varL}}$ respectively, and a length- and power-dependent part ($c^{\mathrm{b,on,varLP}}$ and $c^{\mathrm{b,off,varLP}}$) that is mainly driven by material costs. The operational expenditures of the offshore grid $c^{\mathrm{HVDC,varOM}}$ are modeled as a discounted cashflow dependent on the capital expenditures. In order to enable the expansion of trading capacities ashore, a length-dependent $c^{\mathrm{NTC,varL}}$ and length- and power-dependent $c^{\mathrm{NTC,varLP}}$ cost term for the expansion of net transfer capacities (NTC) is considered. OWP costs consist of power-dependent capital $c^{\mathrm{OWP,varP}}$ expenditures that include costs for the actual wind turbines and wind-park-internal AC-wiring as well as costs for foundation and construction. Furthermore, capital-dependent operational $c^{\mathrm{OWP,varOM}}$ expenditures are considered. 
The applied parameters including costs for primary fuel and CO\textsubscript{2}-emissions and their sources are presented in Table \ref{tab_cost}. 

Scenario data of the pan-European electricity markets for the target year 2030 is based on the ten year network development plan (TYNDP) 2020 \cite{TYNDP.2020} and the federal German network development plan \cite{UbertragungsnetzbetreiberDeutschland.} 2030. For the \mbox{ENTSO-E} region outside Germany, the \textit{Distributed Energy} scenario from the TYNDP is adopted while for Germany data from the federal network development plan (scenario \textit{B}) is implemented. The considered generation capacities and the annual load figures of the neighboring countries of the North Sea are presented in Table \ref{tab_scenario}. Note that in order to model the wind capacity expansion endogenously, the installed offshore wind capacity in the scenario year 2030 has been set to the current (2021) installed capacity. 

\section{Results} \label{sec:results}
\begin{figure}
	\begin{minipage}[c]{.48\textwidth}
		\begin{tcolorbox}[colback=white, sharp corners]
			\begin{subfigure}{\textwidth}
				\includegraphics[width=\textwidth]{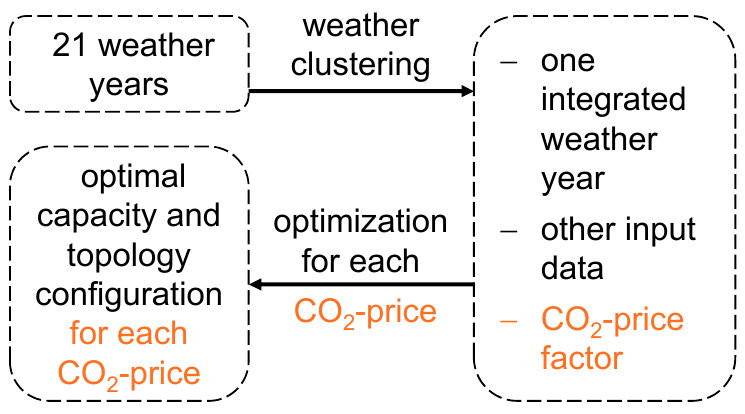}
				\caption{Multi-year approach}
				\label{fig:spap1}
			\end{subfigure}%
			\hfill
			\begin{subfigure}{\textwidth}
				\includegraphics[width=\textwidth]{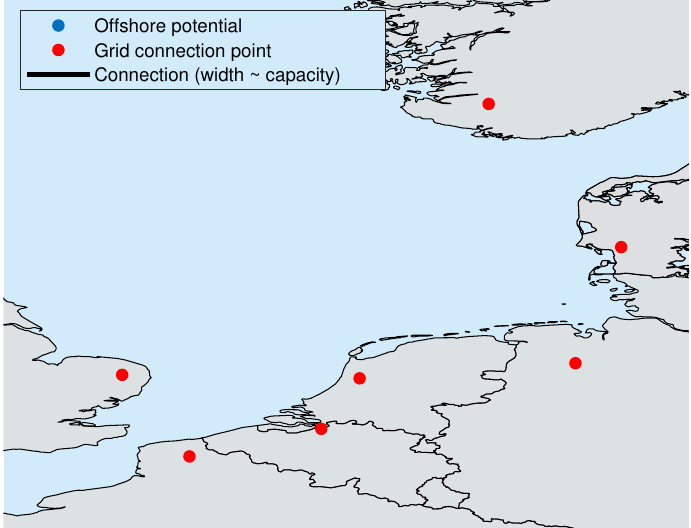}
				\caption{$c^{\mathrm{CO_2}} = 133$ \texteuro/tCO\textsubscript{2}}
				\label{fig:sfig1}
			\end{subfigure}%
			\hfill
			\begin{subfigure}{\textwidth}
				\includegraphics[width=\textwidth]{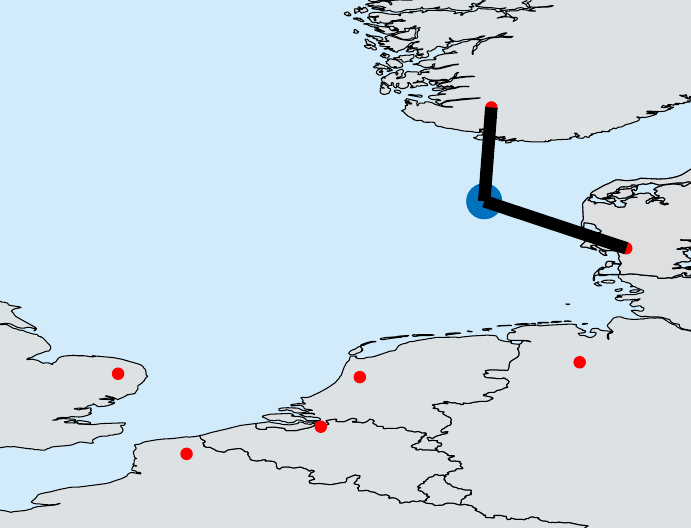}
				\caption{$c^{\mathrm{CO_2}} = 159$ \texteuro/tCO\textsubscript{2}}
				\label{fig:sfig4}
			\end{subfigure}%
			\hfill
			\begin{subfigure}{\textwidth}
				\includegraphics[width=\textwidth]{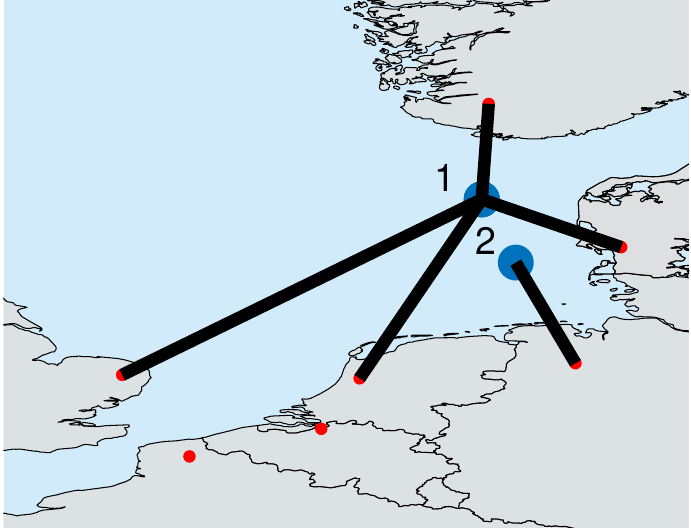}
				\caption{$c^{\mathrm{CO_2}} = 186$ \texteuro/tCO\textsubscript{2}}
				\label{fig:sfig2}
			\end{subfigure}%
		\end{tcolorbox}
		\caption{Evolution of grid topologies considering an increasing CO\textsubscript{2} price using the integrated multi-year weather approach.} 
		\label{fig:figd} 
	\end{minipage}%
	\hfill
		\begin{minipage}[c]{.48\textwidth}
			\begin{tcolorbox}[colback=white, sharp corners] 
				\centering 
				\begin{subfigure}{\textwidth}
					\includegraphics[width=\textwidth]{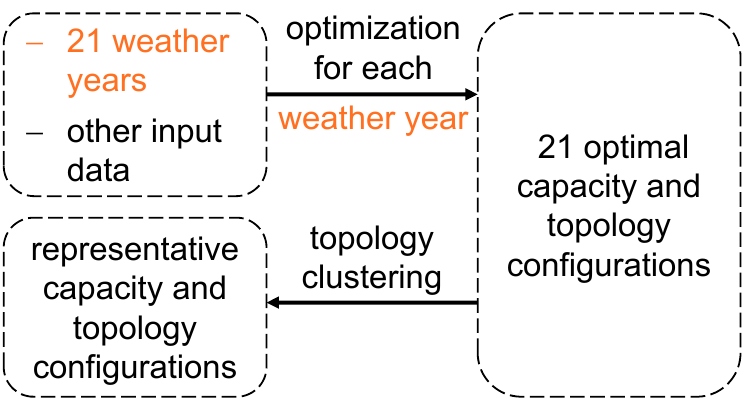}
					\caption{Single-year approach}
					\label{fig:spap2}
				\end{subfigure}%
				\hfill
				\begin{subfigure}{\textwidth}
					\includegraphics[width=\textwidth]{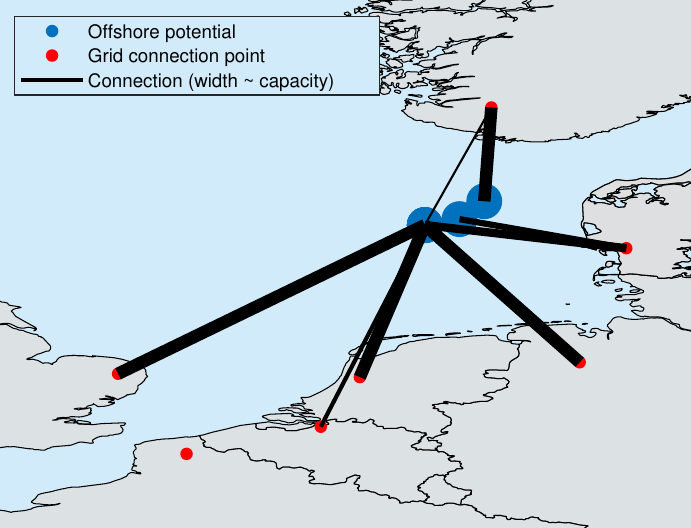}
					\caption{$c^{\mathrm{CO_2}} = 186$ \texteuro/tCO\textsubscript{2}}
					\label{fig:sfig5}
				\end{subfigure}%
				\hfill
				\begin{subfigure}{\textwidth}
					\includegraphics[width=\textwidth]{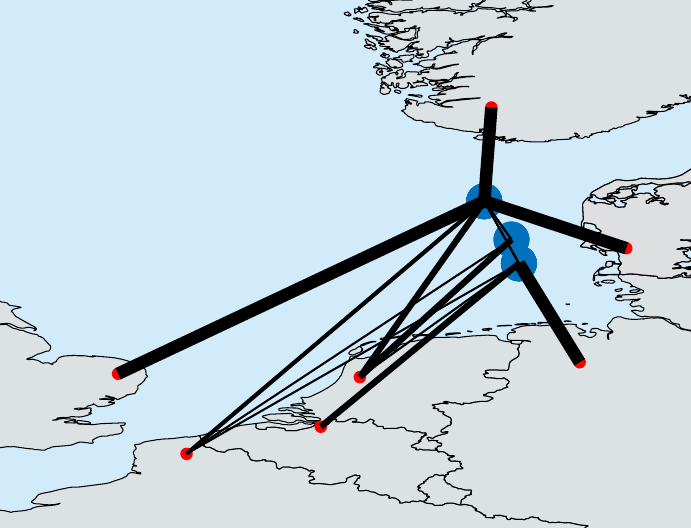}
					\caption{$c^{\mathrm{CO_2}} = 186$ \texteuro/tCO\textsubscript{2}}
					\label{fig:sfig3}
				\end{subfigure}%
				\hfill
				\begin{subfigure}{\textwidth}
					\includegraphics[width=\textwidth]{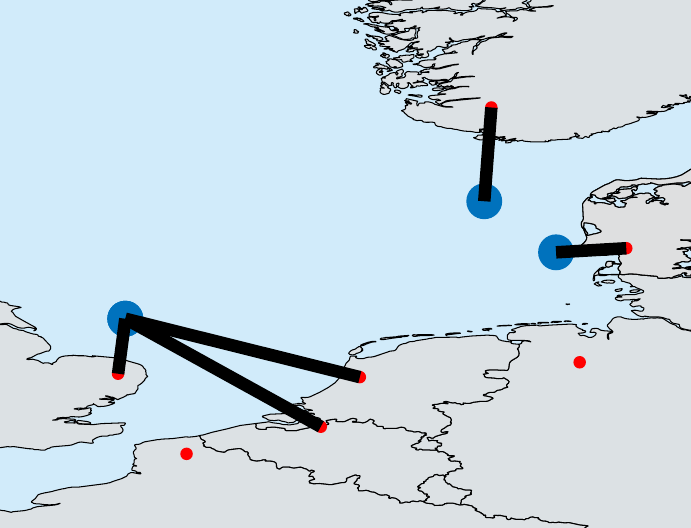}
					\caption{$c^{\mathrm{CO_2}} = 186$ \texteuro/tCO\textsubscript{2}}
					\label{fig:sfig6}
				\end{subfigure}%
			\end{tcolorbox}
			\caption{Different representative topology configurations considering the single-year weather approach for a CO\textsubscript{2} price of 186 \texteuro/ton.} 
			\label{fig:figsd} 
		\end{minipage}%
	\end{figure}

In order to assess the optimal capacity and topology configuration\footnote{The term \textit{optimal capacity and topology configuration} refers to the combination of the economically optimal OWP generation capacity and the corresponding grid capacity and topology.}, two different approaches that differ in terms of incorporating the weather data are implemented, evaluated and compared against. In the first approach, the available weather data are temporally clustered to one representative year. Then, the CO\textsubscript{2} price is gradually increased\footnote{The base price of \texteuro53\,/tCO\textsubscript{2} is multiplied by a factor that is incremented by 0.5 in each scenario.} to evaluate the effect of increasing variable costs on the offshore generation and grid capacity expansion problem (Figure \ref{fig:figd}). In the second approach, rather than incorporating the available weather data in an integrated manner, a single optimization is carried out for each of the 21 available weather years considering constant prices. This approach yields 21 optimal capacity and topology configurations which are then spatially clustered to three representative candidates (Figure \ref{fig:figsd}). This method is applied to demonstrate the effects of using an integrated weather approach that is designed to comprise of representative data over the assets' lifetime against an investment decision based solely on a single weather year.

The approaches have been implemented in the \textit{MILES} framework which covers the whole ENTSO-E area in hourly resolution. Simulations are carried out using MATLAB ver. R2022a in combination with the toolbox for optimization modeling YALMIP and Gurobi Optimizer 10.0.3 \cite{TheMathWorksInc..2022, Lofberg2004, gurobi}. 
The optimization problem comprises around 3,816,000 continuous and 956 integer (481 binary) free variables after presolving. On an Intel(R) Xeon(R) CPU E5-2640 v4 @ 2.40\,GHz with 32\,GB RAM, solving times ranged from around 11\,h (31,287 gurobi work units, Figure \ref{fig:sfig1}) to 34\,h (78,778 gurobi work units, Figure \ref{fig:sfig2}) using one core. The results have been obtained using a spatial granularity of 25 clusters which guarantees a maximum distance between a single offshore wind turbine and its nearest HVDC terminal of approx. 70\,km. The optimization covers 8 annualized representative weeks (168\,hours) in hourly resolution (see Figure \ref{fig:zeitlichesClustering}).

Figures \ref{fig:sfig1} \ref{fig:sfig4} \ref{fig:sfig2} show the resulting configurations for the integrated weather optimization considering different CO\textsubscript{2} prices. Due to a lack of cost advantages, no offshore wind and grid expansion can be observed up to a CO\textsubscript{2} price of 133\,\texteuro/ton. From there on, offshore wind becomes increasingly lucrative so that for a price of \texteuro159\,/tCO\textsubscript{2}, 12\,GW of additional offshore wind capacity is built and connected to NO\footnote{For readability ISO-3166 country codes are used.} and DKW. For a CO\textsubscript{2}-price of 186\,\texteuro/ton, the offshore wind capacity of OWP 1 is extended to 24\,GW and additional connections are made to UK and NL. Furthermore, a second OWP with a total capacity of 6\,GW in proximity to DE is built and connected. In contrast, Figures \ref{fig:sfig5}, \ref{fig:sfig3}, \ref{fig:sfig6} show the representative candidates of the 21 single (weather) optimizations for a CO\textsubscript{2}-price of 186\footnote{The CO\textsubscript{2}-price of 186\,\texteuro/ton has been chosen in order to ensure ample generation and transmission capacity expansion for the comparison of the two approaches.} \texteuro/ton. The candidates have been determined using standard k-medoids clustering on the resulting topologies represented by their adjacency matrix and their corresponding capacities. The topologies based on single weather years show some common features and also substantial differences. Generally, the topologies are radially structured and an expansion can be observed in the north-east area between NO and DKW. Although, in configuration \ref{fig:sfig6} the OWP capacity is generally more distributed throughout the North Sea, while in \ref{fig:sfig5} and \ref{fig:sfig3} capacity is only built in the north-eastern part. 
As configuration \ref{fig:sfig2} is expected to be the most robust one due to the implemented weather approach, further examinations with regards to its effect on the European electricity markets are carried out and presented in the following. 

\begin{figure}
	\centering
	\includegraphics[width=\textwidth]{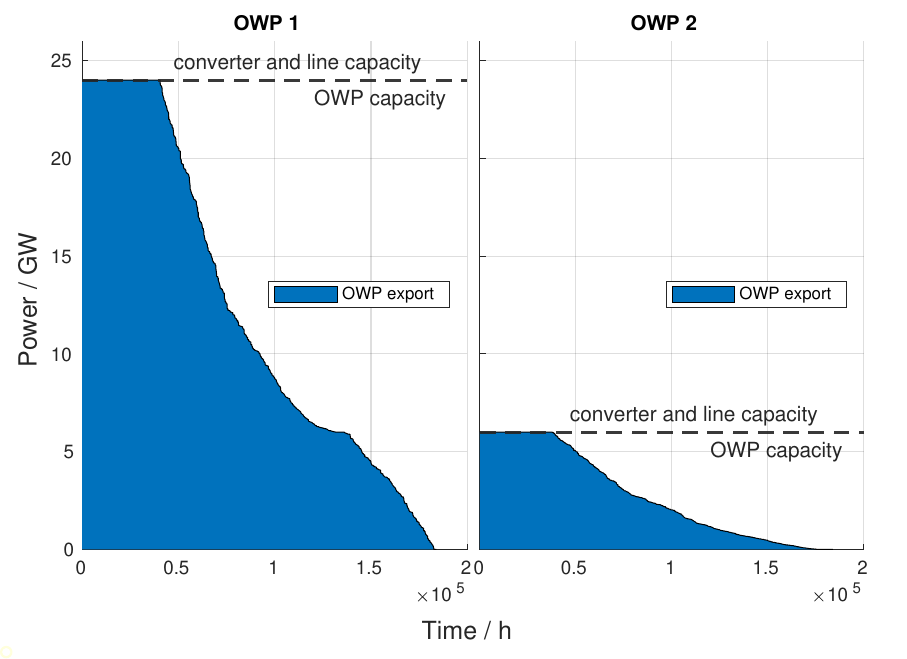}
	\caption{Feed-in duration curves for OWP 1 and 2 including the transit from neighboring countries calculated for all years.} 
	\label{fig:Jahresdauerlinien} 
\end{figure}

Two OWPs - in the following OWP 1 and OWP 2 - are built and connected in configuration \ref{fig:sfig2}. OWP 1 is connected with a radial structure to UK, NO, DKW and NL while OWP 2 is only connected to DE. The total available OWP capacity for clusters 1 and 2 account to 24.42\,GW and 11.77\,GW out of which 24\,GW and 6\,GW respectively have been built\footnote{The scenario consider existing offshore wind capacity of 31\,GW.} (Figure \ref{fig:Jahresdauerlinien}). To transport the electricity ashore, converter and line capacities have to be built, which in principle can be built independently from each other. Nevertheless, the line and converter capacities have been determined to have the same value as the installed OWP capacities for both OWPs, see Figure \ref{fig:Jahresdauerlinien}. The plateaus result from the technical cut-out speed of wind turbines which are considered in the calculation of the capacity factors. The results exhibit no additional market-based curtailment of wind power. In order to prevent grid congestion in the continental European AC-transmission grid, a maximum landing power of 6 GW per bidding zone from all connected OWPs has been set as a restriction which is a binding constraint for both OWPs. 
It is worth noting that both chosen OWP sites exhibit large amounts of full load hours ranking them in the 95\,\% quantil of all available sites indicating a strong sensitivity of the results towards OWP full load hours. This is mainly due to two effects, (i) no regionally differentiated OWP erection costs are considered, thus, it is highly beneficial to exploit the available full load hours. (ii) In the current cost composition, the construction of OWPs including their foundation, internal AC-wiring and the actual wind turbines, constitute a major portion of the total CAPEX of the expansion problem further incentivizing best possible utilization of full load hours. 

\begin{figure}
	\centering
	\includegraphics[width=\textwidth]{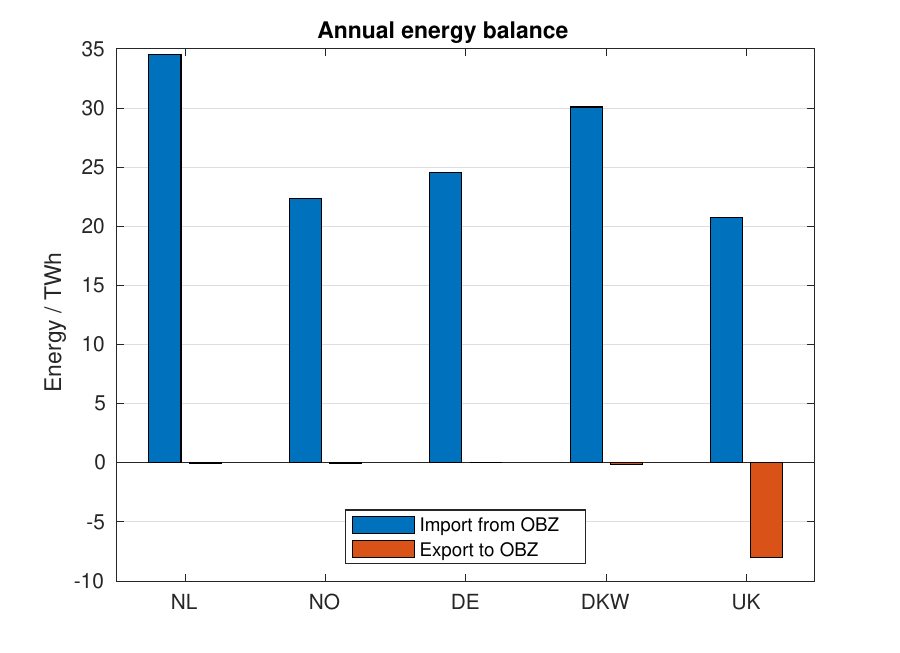}
	\caption{Annual energy balance of interconnected bidding zones with OWPs.} 
	\label{fig:TradeBalances} 
\end{figure}

Figure \ref{fig:TradeBalances} shows the annual trade balances between the built OWPs and the (inter)connected bidding zones. The large numbers of imports from OWPs in contrast to small amounts of export to OWPs indicate that the offshore grid is mainly utilized to transport the wind power ashore and only to a lesser extent to foster international electricity trade. The small export figures can partially be explained by the built-in possibility to built additional interconnectors in the continental European AC-transmission grid. Thus, in the optimization it is cost-effective to expand trading capacities by building additional interconnectors ashore rather than expanding the offshore grid for  electricity trading. 
An optimization run carried out without the possibility to build continental interconnectors showed a substantial offshore grid expansion and export numbers to OWPs, especially from the French bidding zone. 
		
Next, evaluations on the resulting consumer and producer surplus of affected bidding zones are carried out. First, the resulting market clearing prices (MCPs) $\pi_{m,t}$ of each bidding zone $m$ have been determined by a downstream linear optimization in which the results of the topology and capacity optimization are considered as exogenous inputs. This allows to obtain the hourly MCPs as the dual variables of the load coverage constraints of the optimization problem \cite{Weber.2022}. By comparing the MCPs of a simulation run without (\textit{Ref}) and with (\textit{Off}) the OWPs, the delta values for the consumer surplus $CS$ and producer surplus $PS$ can be obtained using \eqref{eq:Delta_Konsumentenrente} - \eqref{eq:Delta_Produzentenrente}. To allow an adequate comparison between the two cases, the optimzed intercontinental NTC values from the \textit{Off} case have also been applied to the \textit{Ref} case. 

\begin{equation} \label{eq:Delta_Konsumentenrente} 
	\begin{split}
		\Delta CS_m = \sum_{t \in \mathcal{M}_{t}} ( P_{m,t}^\mathrm{Dem, Ref} \cdot \pi_{m,t}^\mathrm{Ref} )  \\
		- \sum_{t \in \mathcal{M}_{t}} ( P_{m,t}^\mathrm{Dem, Off} \cdot  \pi_{m,t}^\mathrm{Off} )
	\end{split}
\end{equation}

\begin{equation} \label{eq:Delta_Produzentenrente} 
	\begin{split}
		\Delta PS_m = \sum_{t \in \mathcal{M}_{t}} \sum_{k \in \mathcal{M}_{\mathrm{Supply,m}}} P_{k,t}^\mathrm{Off} (\pi_{m,t}^\mathrm{Off} - c_{k,t}) \\
		- \sum_{t \in \mathcal{M}_{t}} \sum_{k \in \mathcal{M}_{\mathrm{Supply,m}}} P_{k,t}^\mathrm{Ref} (\pi_{m,t}^\mathrm{Ref} - c_{k,t})
	\end{split}
\end{equation}

\begin{figure}
	\centering
	\includegraphics[width=\textwidth]{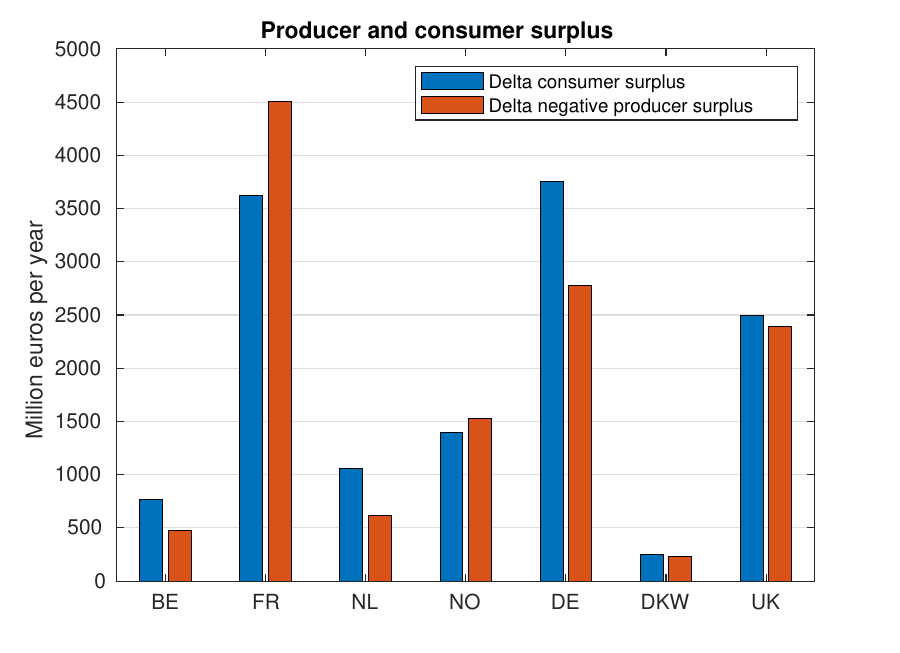}
	\caption{Delta figures for the producer and consumer surplus of affected bidding zones.} 
	\label{fig:KonsumentenProduzentenRente} 
\end{figure}

The feed-in from OWPs lead to a displacement of more expensive energy sources which results in decreased market prices.
The delta comparison of the producer and consumer surplus - depicted in Figure \ref{fig:KonsumentenProduzentenRente} - shows that consumers can capitalize (strongly) from decreased MCPs. On the other hand, certain producers are displaced and on average all producers have to sell their electricity at lower prices resulting in negative producer surpluses. In principle, the economic welfare is the sum of both producer and consumer surplus [quelle]. Thus, if the increase in consumer surplus exceeds the loss of producer surplus, the overall economic welfare\footnote{The economic welfare is a complex matter that is influenced by many other factors as well. Thus, the figures presented can only serve as an indicator and have to be put into perspective amidst their respective national economy and population size which requires successive detail analysis out of the scope of this paper.} of the bidding zone increases. Two bidding zones which are influenced heavily in absolute terms by the feed-in of offshore wind power are FR and DE. While in both bidding zones consumers profit from decreasing market prices, the losses of producers in France exceed the gains of consumers while this is not the case in Germany. The effects in France are the result of a displacement of highly-profitable french nuclear power by the offshore wind in Europe. The effect is less pronounced in Germany, where mainly less-profitable, gas-fired power plants are displaced.

\section{Discussion and Conclusion} \label{sec:discussion}
One contribution of this study has been the endogenous modeling of the OWP capacity as well as the grid topology and capacity considering a high spatial and temporal level of detail. In this regard, the optimization provides robust results and the findings are in line with other research on this topic. The findings reaffirm that the profits and losses of an offshore grid and wind power expansion are not distributed equally across European countries. On the one hand, consumers benefit from decreasing market prices and the operators of the newly built offshore wind turbines and grid connections can expect sufficient returns to finance their investments. On the other hand, operators of fossil-fuel and nuclear power plants lose market shares which can lead to imputed welfare loses in some bidding zones.

In this study, a carbon price of about 200\,\texteuro/tCO\textsubscript{2} has lead to a substantial expansion of offshore wind power in the North Sea. In consideration of the potentially much higher social cost of carbon, this finding stresses the importance of offshore wind parks and the associated grid infrastructure as an outstanding project of common European interest. 

One major sensitivity of the optimization has been the continental transmission grid which has been mapped using a NTC approach. It was necessary to introduce additional endogenous variables to allow for a possible expansion of continental grid interconnections to increase trading capacities. Otherwise, the offshore expansion was strongly driven by export volumes from the French and British bidding zone which is deemed unlikely since the expansion on the mainland seems the economically favorable option.

The optimal capacity and topology configurations exhibit strong sensitivity towards to underyling cost assumptions. Especially the construction of offshore HVDC connections are quite heterogeneous and historical cost data is scarce and exhibits large spreads. In general, the compilation of a consistent cost dataset for the scenario year 2030 including costs for offshore wind turbines, their internal (AC-)wiring as well as primary energy and CO\textsubscript{2}-emission prices is a complex undertaking.  As indicated in the result section, especially the interaction between OWP full load hours, their construction costs and their embedding in the overall cost composition demands a future closer examination in order to obtain more robust results. Therefore, it would be desirable to apply stochastic cost estimations considering cost degressions, regionally different erection costs for OWPs and converter substations as well as different primary fuel price paths.

It is worth noting that no contingency or stability aspects have been considered in this study. Thus, for a holistic solution further detailed analysis of the proposed HVDC grid infrastructure and its interaction with the continental AC grid with special emphasis on technical feasibility and robustness should be carried out next. 

By applying an integrated multi-year weather approach, robust results with regard to the weather influence are obtained. The comparison against the common single-weather year approach often used in the literature highlights the strong influence the weather has on the resulting optimal capacity and topology configuration and stresses the importance of considering weather uncertainty in energy system design. Here, a modified k-medoids clustering algorithm has been applied to determine the representative weeks out of the total 21-year data. One peculiarity of the clustering algorithm lies in its non-deterministic behavior meaning that multiple runs do not always yield the same results. Indeed, rerunning the multi-year weather approach a number of times propagated in negligible variations of the resulting optimal capacity and topology configuration. Nevertheless, our focus in this study was to first highlight the importance of weather uncertainty. In order to obtain results as robust as possible in the future, adequate methods should be developed and applied to minimize the stochastic behavior of the temporal clustering algorithm. Furthermore, the approach should in the future be modified to not only include historical weather data but also extrapolated synthetic data to quantify the impact of climate change on the design of energy systems. 

In order to keep the problem complexity in check, the optimization required a balancing between the number of endogenous variables and the spatio-temporal level of detail. In this study, we tried to maximize the number of endogenous variables and hence applied methods for complexity reduction in both the spatial and temporal domain. Due to the outstanding importance of offshore wind power for the future European energy system, further research considering an even higher granularity is desirable. Therefore, the presented approaches could serve as starting point for the further development of classical optimization techniques, heuristic and meta-heuristic approaches and methods for problem complexity reduction.

\section*{Acknowledgments}

The authors gratefully acknowledge the computing time provided on the Linux HPC cluster at Technical University
Dortmund (LiDO3), partially funded in the course of the Large-Scale Equipment Initiative by the German Research Foundation (DFG) as project 271512359. This research was partly funded by the German Federal Ministry for Economic Affairs and Climate Action within the project GreenVEgaS (funding code: 03EI1009A).

\appendix
	
\section{Model}\label{sec:model}
The study has been carried out using a modified version of the pan-European electricity market model MILES \cite{MILES} that also includes the offshore expansion problem. In the following, a reduced set of equations with focus on the implemented generation and transmission capacity expansion problem is presented. The objective of the optimization problem is to minimize the sum of the operational expenditures of the pan-European power plant dispatch $OpEx^{\text{Dispatch}}$ considering the annualized costs for the expansion of offshore generation $AC^{\text{OWP}}$ and transmission $AC^{\text{HVDC}}$ capacities \eqref{eq:objective}.	
	
\begin{equation} \label{eq:objective} 
	min(OpEx^{\text{Dispatch}} + AC^{\text{OWP}} + AC^{\text{HVDC}})
\end{equation}

The operational expenditures result from the pan-European dispatch of conventional power plants to satisfy the electrical demands and control reserve needs in each bidding zone. They comprise of the power feed-in $P_{k,t,z}$ of thermal plant $k$ in time step $t$ in week cluster $z$ multiplied by a cost factor $c_{k,t,z}$ that covers fuel and CO\textsubscript{2}-emission costs as well as additional costs for operation and maintenance. The total costs for each typical week are then multiplied by a weighting factor $N_z$ that describes the occurrence of the typical week over the whole year \eqref{eq:TC_dispatch}. 

\begin{equation} \label{eq:TC_dispatch} 
	\begin{split}
		OpEx^{\text{Dispatch}} &= \sum_{z \in \mathcal{M}_{z}} (\sum_{t \in \mathcal{M}_{t}} \sum_{k \in \mathcal{M}_{\text{Units}}} P_{k,t,z} \cdot c_{k,t,z}) \cdot N_z 
	\end{split}
\end{equation}

Equation \eqref{eq:AC_OWP} describes the annualized cost (AC) for the construction and operation of OWPs in the North Sea. It comprises the installed capacity $P^{\text{max, OWP}}_f$ of offshore wind park $f$ multiplied by power-dependent investment costs $c^{\mathrm{OWP,varP}}$ which cover expenses for wind turbines, park-internal AC-wiring and the foundation and construction. In order to obtain annualized costs the capital expenditures are multiplied with a capital recovery factor (CRF) $\text{crf}^{\text{OWP}}$ and recurring costs for operation and maintenance (om) $\text{om}^{\text{OWP}}$ which are modeled as a fraction of the investment costs. 

\begin{equation} \label{eq:AC_OWP} 
	\begin{split}
		AC^{\text{OWP}} &= \sum_{f \in \mathcal{M}_{f}} P^{\text{max, OWP}}_f \cdot c^{\mathrm{OWP,varP}}_f \cdot (\text{crf}^{\text{OWP}} + \text{om}^{\text{OWP}}) 
	\end{split}
\end{equation}

The annualized costs for the offshore grid are split up between (i) routing and wiring costs for the dc transmission and (ii) converter and auxiliary power electronic costs \eqref{eq:AC_HVDC}. The individual cost components will be explained in detail further below after the description of the OWP modeling. 

\begin{equation} \label{eq:AC_HVDC} 
	\begin{split}
		AC^{\text{HVDC}} &= AC^{\mathrm{c, fix}} + AC^{\mathrm{c,varP,ac-dc}} +  AC^{\mathrm{c,varP,dc-dc}} \\ 
		&+ AC^{\mathrm{b, fix}} + AC^{\mathrm{b,varL}} + AC^{\mathrm{b,varLP}} \nonumber
	\end{split}
\end{equation}

Each OWP is modeled as its own offshore bidding zone (OBZ) and integrated via an NTC approach assuming fully controllable power flows along the HVDC connections. The maximum feed-in from an OWP $P^{\text{OWP}}_{f,t,z}$ is restricted by the optimal installed capacity  $P^{\text{max, OWP}}_f$ multiplied by a time-dependent capacity factor $CF^{\text{OWP}}_{f,t,z}$ taking into account the current weather situation and system curve of the OWP \eqref{eq:Production_OBZ}. The inequality allows for power curtailment in case of non-integratable wind energy. The installed capacity per wind park has to be within the predefined limits $P^{\text{maxAvailable, OWP}}_f$ mainly driven by the available area \eqref{eq:Production_OBZ_maxAvailable}. 

\begin{subequations}
	\begin{align}
		0 \leq P^{\text{OWP}}_{f,t,z} \leq CF^{\text{OWP}}_{f,t,z} \cdot P^{\text{max, OWP}}_f \label{eq:Production_OBZ} \\
		P^{\text{max, OWP}}_f \leq P^{\text{maxAvailable, OWP}}_f \label{eq:Production_OBZ_maxAvailable} \\
		\quad \forall \: f \in \mathcal{M}_{OBZ}, \forall \: t \in \mathcal{M}_{t}, \forall \: z \in \mathcal{M}_{z} \nonumber
	\end{align}
\end{subequations}

The power balance of each OBZ is equal to the power feed-in from the OWP minus the net export position $P^{\text{NEx}}_{m,t,z}$ \eqref{eq:LeistungsGG}. The net export position of an OBZ can be described as the sum of the exports from one OBZ to all other bidding zones minus the sum of imports from all other bidding zones \eqref{eq:P_netto}. The sets $\mathcal{M}_{m}$, $\mathcal{M}_{OBZ}$ and $\mathcal{M}_{ML}$ describe the elements of (i) all bidding zones, (ii) all OBZs and (iii) all mainland (ML) bidding zones with a permissible connection to an OBZ, respectively.

\begin{equation} \label{eq:LeistungsGG} 
	\begin{split}
		P^{\text{OWP}}_{f,t,z} - P^{\text{NEx}}_{m,t,z} = 0 \\
		\quad \forall \: f \in \mathcal{M}_{OBZ}, \forall \: m \in \mathcal{M}_{OBZ}, \forall \: t \in \mathcal{M}_{t}, \forall \: z \in \mathcal{M}_{z}
	\end{split}
\end{equation}

\begin{equation} \label{eq:P_netto} 
	\begin{split}
		P^{\text{NEx}}_{m,t,z} = \sum_{m2 \in \mathcal{M}_{m}} P_{m1,m2,t,z} - \sum_{m2 \in \mathcal{M}_{m}} P_{m2,m1,t,z} \\
		\quad \forall \: m1 \in \mathcal{M}_{mOBZ}, \forall \: m2 \in \mathcal{M}_{m} \forall \: t \in \mathcal{M}_{t}, \forall \: z \in \mathcal{M}_{z}
	\end{split}
\end{equation}

The power flow $P_{m1,m2,t,z}$ from one market area to another is restricted by an integer variable $NTC^{\text{Step}}_{m1,m2}$ counting the number of connections with power $s$ \eqref{eq:Pexport_hin_OBZ} and \eqref{eq:Pexport_hin_OBZ}. Furthermore, an adjacency matrix $A_{m2,m1}$ is implemented to account for fixed costs using a big-M approach \eqref{eq:Adjazenz_bigM}. Since the power flow can be directional, the NTC matrix is symmetrical \eqref{eq:NTC_max_hin_zurueck}. Lastly, the maximum permissible power flow between mainland connections and OBZs is restricted to account for the limited hosting capacity of the continental AC grid \eqref{eq:MaxLeistungLand}.

\begin{subequations}
	\begin{align}  
		P_{m1,m2,t,z} \leq (\text{NTC}^{\text{Step}}_{m1,m2} \cdot s) \cdot A_{m1,m2} \label{eq:Pexport_hin_OBZ} \\
		\quad \forall \: m1 \in \mathcal{M}_{OBZ}, \forall \: m2 \in \mathcal{M}_{m}, \forall \: t \in \mathcal{M}_{t}, \forall \: z \in \mathcal{M}_{z} \nonumber \\
		P_{m2,m1,t,z} \leq (\text{NTC}^{\text{Step}}_{m2,m1} \cdot s) \cdot A_{m2,m1} \label{eq:Pexport_zurueck_OBZ} \\
		\quad \forall \: m1 \in \mathcal{M}_{m}, \forall \: m2 \in \mathcal{M}_{OBZ}, \forall \: t \in \mathcal{M}_{t}, \forall \: z \in \mathcal{M}_{z} \nonumber
	\end{align}
\end{subequations}

\begin{equation} \label{eq:Adjazenz_bigM} 
	\text{NTC}^{\text{Step}}_{m1,m2} \leq M \cdot A_{m2,m1}  
\end{equation}

\begin{equation} \label{eq:NTC_max_hin_zurueck} 
	\begin{split}
		\text{NTC}^{\text{Step}}_{m1,m2} = \text{NTC}^{\text{Step}}_{m2,m1} \\
		\quad \forall \: m1 \in \mathcal{M}_{OBZ}, \forall \: m2 \in \mathcal{M}_{m}, \forall \: t \in \mathcal{M}_{t}, \forall \: z \in \mathcal{M}_{z}
	\end{split}
\end{equation}

\begin{subequations} \label{eq:MaxLeistungLand} 
	\begin{align} 
		\sum_{m1 \in \mathcal{M}_{OBZ}} P_{m1,m2,t,z} \leq P^{\mathrm{ML,max}} \\
		\sum_{m1 \in \mathcal{M}_{OBZ}} P_{m2,m1,t,z} \leq P^{\mathrm{ML,max}} \\
		\forall \: m2 \in \mathcal{M}_{ML}, \forall \: t \in \mathcal{M}_{t}, \forall \: z \in \mathcal{M}_{z} \nonumber
	\end{align} 
\end{subequations}

Next, the cost terms of the expansion problem are described. $AC^{\mathrm{b, fix}}$ represents fixed routing costs that are independent of the actual length of the corridor \eqref{eq:Kabel_fix}. The length-dependent variable costs $AC^{\mathrm{b,varL}}$ are determined by the distance $d_{m1,m2}$ between two bidding zones, the share of covered land and sea distance $a_{m1,m2}^{\mathrm{on}}$ and $a_{m1,m2}^{\mathrm{off}}$ and the corresponding cost terms $c^{\mathrm{b,on,varL}}$ and $c^{\mathrm{b,off,varL}}$ multiplied by the capital recovery factor of an HVDC grid $(crf^{\mathrm{HVDC}}$ \eqref{eq:Kabel_varLänge} and \eqref{eq:Kabel_varLängeKosten}. Analogous, a length- and power-dependent part/share $AC^{\mathrm{b,varLP}}$, mainly driven by material costs, is considered \eqref{eq:Kabel_varLängeLeistung} and \eqref{eq:Kabel_varLängeLeistungKosten}. Since the adjacency matrix is symmetrical, but the costs arise only once, the cost terms are halved when appropriate.

\begin{equation} \label{eq:Kabel_fix} 
	AC^{\mathrm{b, fix}} = (\sum_{m2 \in \mathcal{M}_{m}} \sum_{m1 \in \mathcal{M}_{m}} A_{m2,m1} \cdot \frac{c^{\mathrm{b,fix}}}{2}) \cdot (crf^{\mathrm{HVDC}})
\end{equation}

\begin{equation} \label{eq:Kabel_varLänge} 
	c_{m1,m2}^{\mathrm{b,varL}} = d_{m1,m2} \cdot a_{m1,m2}^{\mathrm{on}} \cdot c^{\mathrm{b,on,varL}} + d_{m1,m2} \cdot a_{m1,m2}^{\mathrm{off}} \cdot c^{\mathrm{b,off,varL}}
\end{equation}

\begin{equation} \label{eq:Kabel_varLängeKosten} 
	AC^{\mathrm{b,varL}} =  (\sum_{m2 \in \mathcal{M}_{m}} \sum_{m1 \in \mathcal{M}_{m}} A_{m2,m1} \cdot \frac{c_{m1,m2}^{\mathrm{b,varL}}}{2}) \cdot (crf^{\mathrm{HVDC}})
\end{equation}

\begin{equation} \label{eq:Kabel_varLängeLeistung} 
	c_{m1,m2}^{\mathrm{b,varLP}} = d_{m1,m2} \cdot a_{m1,m2}^{\mathrm{on}} \cdot c^{\mathrm{b,on,varLP}} + d_{m1,m2} \cdot a_{m1,m2}^{\mathrm{off}} \cdot c^{\mathrm{b,off,varLP}}
\end{equation}

\begin{equation} \label{eq:Kabel_varLängeLeistungKosten} 
	AC^{\mathrm{b,varLP}} =  (\sum_{m2 \in \mathcal{M}_{m}} \sum_{m1 \in \mathcal{M}_{m}} (\text{NTC}^{\text{Step}}_{m2,m1} \cdot s) \cdot \frac{c_{m1,m2}^{\mathrm{b,varLP}}}{2}) \cdot (crf^{\mathrm{HVDC}} + c^{\mathrm{HVDC,varOM}})
\end{equation}

Converter costs are modeled to comprise a fixed part $AC^{\mathrm{c, fix}}$ for the foundation and erection of sea platforms \eqref{eq:Konverter_fix} and two power-dependent parts taking into account if the connection is established between AC-DC $AC^{\mathrm{c,varP,ac-dc}}$ \eqref{eq:KonverLeistung_ac-dc} or DC-DC $AC^{\mathrm{c,varP,dc-dc}}$ \eqref{eq:KonverLeistung_multiterminal}. The converter power of an AC-DC connection is designed to handle the maximum occurring power flow considering both directions OBZ $\rightarrow$ ML $P^{\mathrm{on,max}}_{m2}$ \eqref{eq:PmaxLand} and ML $\rightarrow$ OBZ $P^{\mathrm{off,max}}_{m1}$ \eqref{eq:PmaxSee}. In case of a multiterminal DC-DC connection, the annualized costs are mainly driven by direct current circuit breakers \eqref{eq:KonverLeistung_multiterminal}.

\begin{equation} \label{eq:Konverter_fix} 
	\begin{split}
		AC^{\mathrm{c, fix}} =(\sum_{m2 \in \mathcal{M}_{m}} \sum_{m1 \in \mathcal{M}_{OBZ}} A_{m1,m2} \cdot \frac{c^{\mathrm{c,fix}}}{2}) \cdot (crf^{\mathrm{HVDC}} + c^{\mathrm{HVDC,varOM}})
	\end{split}
\end{equation}

\begin{subequations}
	\begin{align}  
		\sum_{m1 \in \mathcal{M}_{OBZ}} P_{m1,m2,t,z} \leq P^{\mathrm{on,max}}_{m2} \label{eq:PmaxLand} \\
		\quad \forall \: m2 \in \mathcal{M}_{ML}, \forall \: t \in \mathcal{M}_{t}, \forall \: z \in \mathcal{M}_{z} \nonumber \\
		\sum_{m2 \in \mathcal{M}_{ML}} P_{m2,m1,t,z} \le P^{\mathrm{off,max}}_{m1}  \label{eq:PmaxSee} \\
		\quad \forall \: m1 \in \mathcal{M}_{OBZ}, \forall \: t \in \mathcal{M}_{t}, \forall \: z \in \mathcal{M}_{z} \nonumber
	\end{align}
\end{subequations}

\begin{equation} \label{eq:KonverLeistung_ac-dc} 
	\begin{split}
		AC^{\mathrm{c,varP,ac-dc}} = ( (\sum_{m2 \in \mathcal{M}_{ML}} P^{\mathrm{on,max}}_{m2} \cdot \frac{c^{\mathrm{c,varP,ac-dc}}}{2}) + \\ (\sum_{m2 \in \mathcal{M}_{OBZ}} P^{\mathrm{off,max}}_{m1} 
		\cdot \frac{c^{\mathrm{c,varP,ac-dc}}}{2}) ) \\ \cdot (crf^{\mathrm{HVDC}} + c^{\mathrm{HVDC,varOM}})
	\end{split}
\end{equation}

\begin{equation} \label{eq:KonverLeistung_multiterminal} 
	\begin{split}
		AC^{\mathrm{c,varP,dc-dc}} = (\sum_{m2 \in \mathcal{M}_{OBZ}} \sum_{m1 \in \mathcal{M}_{OBZ}}(\text{NTC}^{\text{Step}}_{m1,m2} \cdot s) \cdot \frac{c^{\mathrm{c,varP,dc-dc}}}{2}) \\ \cdot (crf^{\mathrm{HVDC}} + c^{\mathrm{HVDC,varOM}})
	\end{split}
\end{equation}

\section{Data Tables}\label{secA2}
\begin{table}[h]
	\begin{center}
		\begin{minipage}{\textwidth}
			\caption{Cost and model parameters}\label{tab_cost}
			\begin{tabular*}{\textwidth}{@{\extracolsep{\fill}}llrr@{\extracolsep{\fill}}}
				\toprule%
				Parameter & Value & Unit & Reference \\
				\midrule
				$c^{\mathrm{b,fix}}$ & 0.7 & M \texteuro & \cite{Hartel.2017} \\
				$c^{\mathrm{b,on,varL}}$ & 4.5 & M \texteuro / km & based on \cite{NEP.2035} \\
				$c^{\mathrm{b,off,varL}}$ & 2 & M \texteuro / km & based on \cite{NEP.2035} \\
				$c^{\mathrm{b,on,varLP}}$ & 1 & M \texteuro / km GW & based on \cite{NEP.2035} \\
				$c^{\mathrm{b,off,varLP}}$ & 1 & M \texteuro / km GW & based on \cite{NEP.2035} \\ [5pt]
				
				$c^{\mathrm{c,fix}}$ & 30 & M \texteuro & based on \cite{Hartel.2017} \\ 
				$c^{\mathrm{c,varP,ac-dc}}$ & 750 & M \texteuro / GW & \cite{NEP.2035} \\
				$c^{\mathrm{c,varP,dc-dc}}$ & 125 & M \texteuro / GW & \cite{NEP.2035, Nieradzinska.2016, cigre.2013} \\
				$c^{\mathrm{HVDC,varOM}}$ & 1 & \% / CAPEX p.a & based on \cite{dena.2014} \\ [5pt]
				
				$c^{\mathrm{NTC,varL}}$ & 2 & M \texteuro / km & own assumption \\
				$c^{\mathrm{NTC,varLP}}$ & 1 & M \texteuro / km GW & own assumption \\ [5pt]
				
				$c^{\mathrm{OWP,varP}}$ & 2.4 & M \texteuro / MW & based on \cite{BVGAssociates.2023} \\
				$c^{\mathrm{OWP,varOM}}$ & 6.25 & \% / CAPEX p.a & based on \cite{BVGAssociates.2023} \\
				$\rho^{\mathrm{OWP}}$ & 7 & MW / km & based on \cite{AndrevanKuijk.2019, BundesamtfurSeeschifffahrtundHydrographie.2022} \\ 
				$n^{\mathrm{OWP}}$ & 27 & MW / km & \cite{BVGAssociates.2023} \\
				$n^{\mathrm{HVDC}}$ & 40 & MW / km & based on \cite{dena.2014} \\
				$i^{\mathrm{OWP}}$ & 0.06 & - & \cite{BVGAssociates.2023} \\
				$i^{\mathrm{HVDC}}$ & 0.05 & - & own assumption \\ 
				$\mathrm{crf^{OWP}}$ & 0.0757 & - & calculated \\ 
				$\mathrm{crf^{HVDC}}$ & 0.0583 & - & calculated \\ [5pt] 			
				
				$c^{\mathrm{Nuclear}}$ & 1.7 & \texteuro / $\mathrm{MWh_{th}}$ & \cite{TYNDP.2020} \\
				$c^{\mathrm{Lignite}}$ & 4 & \texteuro / $\mathrm{MWh_{th}}$ & \cite{TYNDP.2020} \\
				$c^{\mathrm{Hard \: coal}}$ & 15.5 & \texteuro / $\mathrm{MWh_{th}}$ & \cite{TYNDP.2020} \\
				$c^{\mathrm{Natural \: gas}}$ & 24.9 & \texteuro / $\mathrm{MWh_{th}}$ & \cite{TYNDP.2020} \\
				$c^{\mathrm{CO_2}}$ & 53 & \texteuro/tCO\textsubscript{2} & \cite{TYNDP.2020} \\  [5pt]
				
				$s$ & 1 & GW & own assumption \\
				$P^{\mathrm{ML,max}}$ & 6 & GW & own assumption \\
				\bottomrule
			\end{tabular*}
		\end{minipage}
	\end{center}
\end{table}

\begin{table}[h]
	\begin{center}
		\begin{minipage}{\textwidth}
			\caption{Scenario parameters 2030 based on \cite{TYNDP.2020}}\label{tab_scenario}
			\begin{tabular*}{\textwidth}{@{\extracolsep{\fill}}llrrrrrrr@{\extracolsep{\fill}}}
				\toprule%
				Capacities & Unit & BE & FR & NL & NO & DE & DKW & UK \\
				\midrule
				\quad Nuclear & GW & 0 & 58.2 & 0.5 & 0 & 0 & 0 & 1.2\\
				\quad Lignite & GW & 0 & 0 & 0 & 0 & 9.3 & 0 & 0 \\
				\quad Hard coal & GW & 0.6 & 0 & 3.4 & 0 & 9.8 & 0.8 & 3.7 \\
				\quad Natural gas & GW & 8.7 & 7.2 & 9.3 & 0 & 35.2 & 1 & 38.7 \\ 
				\quad Mixed fuels & GW & 1.3 & 6.5 & 3.8 & 0.3 & 4.1 & 0.5 & 7.4 \\ 
				\quad Hydropower & GW & 1.5 & 25.3 & 0.1 & 36.1 & 17.2 & 0 & 6.0 \\
				\quad Onshore wind & GW & 5.9 & 44.0 & 8.3 & 8.2 & 81.5 & 4.6 & 25.5 \\ 
				\quad Offshore wind & GW & 2.3 & 3.0 & 3.0 & 0 & 7.8 & 2.3 & 12.5 \\ 
				\quad Photovoltaics & GW & 13.9 & 42.2 & 15.5 & 2.0 & 91.3 & 3.1 & 35.4 \\ [5pt]
				Load & TWh & 101.1 & 483.3 & 137.7 & 150.9 & 534 & 31.5 & 370.3 \\
				\bottomrule
			\end{tabular*}
		\end{minipage}
	\end{center}
	Note: The installed offshore wind power capacities reflect the situation of 2021. Other fossil fuels not extra listed are subsumed under mixed fuels.
\end{table}
		
\printbibliography
			
\end{document}